\documentclass[twocolumn,pra,aps]{revtex4-1}

\usepackage{color}
\usepackage{epsfig}
\usepackage{latexsym}
\usepackage{amssymb}
\usepackage{amsmath}

\usepackage{algorithm}
\usepackage{algorithmic}
\usepackage{wrapfig}
\usepackage[apple mac]{inputenc}
\usepackage[english]{babel}
\usepackage{times}
\usepackage{latexsym}
\usepackage{fancyhdr}
\usepackage{verbatim}
\usepackage{tabularx}
\usepackage{epsfig}
\usepackage{amsmath}
\usepackage{amssymb}
\usepackage{graphicx}
\usepackage{wasysym}

\usepackage{yfonts}

\newcommand{\qed}{\hspace*{\fill}$\square$}

\newcommand{\be}{\begin{equation}}
\newcommand{\ee}{\end{equation}}
\begin{document}

\title{Is the Majorana bound state a gravitationally neutral system?}
\author{Miguel A. Martin-Delgado}
\affiliation{Departamento de F\'{\i}sica Te\'orica, Universidad Complutense, 28040 Madrid, Spain.\\
CCS-Center for Computational Simulation, Campus de Montegancedo UPM, 28660 Boadilla del Monte, Madrid, Spain.}

\begin{abstract} 
An argument is provided that confronts the strong equivalence principle (SEP) of general relativity with the superposition principle (SP) of quantum mechanics. The result has implications on the possibility of synthesizing Majorana bound states in 1D systems or the possible violation of the strong equivalence principle by quantum effects. The argument is formulated  by inducing a very special fine tuning of coupling constants in the Kitaev Hamiltonian representing 1D topological superconductors.
\end{abstract}

\maketitle

%%%%%%%%%%%%%%%%%%%%%%%%%%%%%%%%%%%%%%
%%%%%%%%%%%%%%%%%%%%%%%%%%%%%%%%%%%%%%
\section{Introduction}
\label{sec:intro}
%%%%%%%%%%%%%%%%%%%%%%%%%%%%%%%%%%%%%%
%%%%%%%%%%%%%%%%%%%%%%%%%%%%%%%%%%%%%%

The theory of general relativity \cite{Einstein_relativity,Carroll,Hartle} describes gravitation at macroscopic, astrophysical and cosmological scales but in principle it could be applicable to other scales like the atomic world were it not for the fact that gravity intensity is negligible at those scales.  In fact,  we do not have reliable evidence of gravitation below the micron scale \cite{Tino,Blakemore,Aspelmayer}.  This tension between quantum and gravitation manifests  not only in fundamental physics,  but also in the basic system of units SI  \cite{newSI}.  Despite being the quantum world inaccessible to gravitation in any practical way to date, we can seek to confront both fundamental theories with the aim at obtaining any hint as to what should be assumed in the search for a compatible theory that embraces both of them.

One of the defining properties of Majorana zero modes is being an electrically neutral quasiparticle.  Here  I investigate to what extent they are gravitationally neutral and its implications. The answer to the question in the title depends on different scenarios where the gravitational properties of MBS can be addressed.

%%%%%%%%%%%%%%%%%%%%%%%%%%%%%%%%%%%%%%
%%%%%%%%%%%%%%%%%%%%%%%%%%%%%%%%%%%%%%
\section{Confronting SEP and SP}
\label{sec:confronting}
%%%%%%%%%%%%%%%%%%%%%%%%%%%%%%%%%%%%%%
%%%%%%%%%%%%%%%%%%%%%%%%%%%%%%%%%%%%%%

The SEP is traditionally formulated in two complementary ways:

\noindent {\em  Gravity without weight.} A body falling in a gravitational field ceases to feel gravitation (weight) locally  and the situation is equivalent to having the body in a region of locally zero gravity.

\noindent  {\em  Weight without gravitation.} A body in a laboratory subjected to a local gravity $g$ is equivalent to being in a laboratory with local acceleration  opposite to that gravity,  $a=-g$.   Gravity can be generated locally from acceleration solely.

It is possible to unify both formulations into a single  formulation as follows.

\noindent {\em  Strong Equivalence Principle.} Gravitation is relative.

In the theory of general relativity, gravitation is relative to acceleration meaning that gravitation can be increased or decreased with respect to accelerating systems of reference.  It is not an absolute quantity.  Thus, gravitation must be measured with respect to some accelerated system of reference.  In turn, acceleration becomes universal since it is equivalent to gravitation.

\noindent {\em  Definition:  Neutral gravitational system.} A system with zero gravitational pull or zero gravitation (with respect to any system of reference).

\noindent {\em  Corollary I:} There can not be a gravitationally neutral system.  If that were the case then we could use that system to measure absolute gravity contrary to the SEP.

Compare this situation with the notion of velocity in special relativity: velocity is also a relative concept depending on the inertial system of reference. Thus, there can not be a system  with absolute vanishing velocity (aka aether or at absolute rest). Otherwise,  we could use it to measure absolute velocities contrary to the relativity principle.

A simple example of a neutral gravitational system is one with zero mass and zero energy. Notice that although a falling system does not feel gravitation locally it does have either mass or energy or both, but gravitation cannot be switched off completely or globally. This observation on gravitational neutral systems applies very generally:  for elementary particles,  composite systems,  a system as a whole or to a subsystem of a given system. This has implications for quasiparticles like MBS.

The concept of gravitational neutral system has a predictive power.

\noindent {\em  Corollary II:} systems of negative mass cannot exist.  By reductio ad absurdum,  in that case, we may consider a composite system of two parts,  one point-like particle of mass $ m $ and the other one with $ -m $, totalling zero mass.   Put the whole system into a still box so that they cannot escape.  Altogether, that system would have zero momentum and zero mass yielding a zero total energy system that is gravitationless.  Thus, the initial assumption of a negative mass particle is impossible since it violates the SEP.

Majorana bound states (MBS) in 1D nanowires are excitations of quasiparticles that, under very special circumstances, carry neither energy, mass nor charge. Despite this absence of properties, these very special quasiparticles represent truly quantum states.  As such, they are candidates for a gravitationally neutral system. Let us examine this more closely. The Hamiltonian for a system of spinless fermions representing their hopping $t$,  chemical potential $\mu$ and p-wave pairing $\Delta$ on a one-dimensional open chain with  $L$ sites is 
\begin{align}
H &= -t\sum_{n=1}^{L-1}\left( f^{\dag}_{n+1} f_{n} + \text{h.c.}\right) - \mu \sum_{n=1}^{L} f^{\dag}_{n}f_{n} \\ \nonumber
    &+ \Delta \sum_{n=1}^{L-1} \left(f^{\dag}_{n} f^{\dag}_{n+1} +  \text{h.c.}\right),
\end{align}
where $f^{\dag}_{n}, f_{n}$ are fermionic creation and annhilation operators of  fermionic modes at site $ n $.   They satisfy the canonical anticonmutation relations: $ \{f^{\dag}_{n}, f_{n}\}=\delta_{_{n,m}} $.  At each site of the chain we may have the vacuum state $|0\rangle  $ with no modes and  the excited state $|1\rangle  $ with one mode.

Notice  that this is a non-relativistic Hamiltonian,  thus there is no depedence on the rest energy $ mc^{2} $ of the fermionic particles.  The mass of fermions do appear in  $t$,  as the hopping parameter depends on the intertial mass of the original particles through the overlapping integral,  which for a single band one-dimensional lattice takes the form
\begin{equation}\label{hopping}
t = \int dx \phi^{\ast}(x) \left( -\frac{\hbar^{2}}{2m} \nabla^{2} + U(x)   \right)  \phi(x-R),
\end{equation}
where $R$ is the equal-spacing separating the sites $ x $ of the lattice,  $ \phi(x) $ the atomic orbitals and 
$ U(x) $ the potential of the crystal lattice.
Gravitational sources of mass do not appear in $ H $ due to their negligible effects at the quantum scale.

To construct a topological phase for this system it is advantageous to change the fermionic representation of operators by the Majorana representation defined as
\begin{equation}
f^{\dag}_{n} := \frac{1}{2}(m_{2n-1} + \text{i}m_{2n}), \quad f_{n} :=\frac{1}{2}(m_{2n-1} - \text{i}m_{2n}),
\end{equation}
so that each standard fermion per site is split  in two Majorana modes. Correspondly, the fermion states are represented as a superposition of Majorana states in accordance to the SP.  Thus,  Majorana modes always come in even numbers being confined inside the sites.  They satisfy an anticommutation algebra
different than ordinary fermions: $ \{m_{2n-1},m_{2n}\} = 0 $,   $ m^{2}_{2n-1} = 1 = m^{2}_{2n} $ similar to Pauli matrices.   However,  a deconfinement of Majorna modes at the end points of the chain happens for the special choice of parameters \cite{Kitaev_Majorana}
\begin{equation}\label{fine_tuning}
\Delta = t,  \quad \mu = 0.
\end{equation}
This special regime of parameters is the topological phase represented by the Majorana chain Hamiltonian
\begin{equation}\label{Kitaev_H}
H_{\text{K}} = \text{i} \Delta \sum_{n=1}^{L-1} m_{2n-1}m_{2n}.
\end{equation}
%%%%%%%
%%%%%%%
\begin{figure}[ht]
\begin{center}
 \includegraphics[scale=0.25]{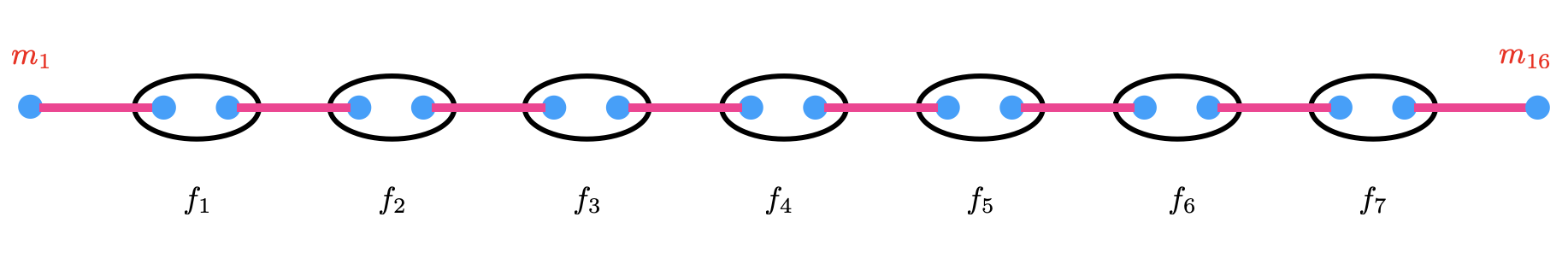}
  \includegraphics[scale=0.4]{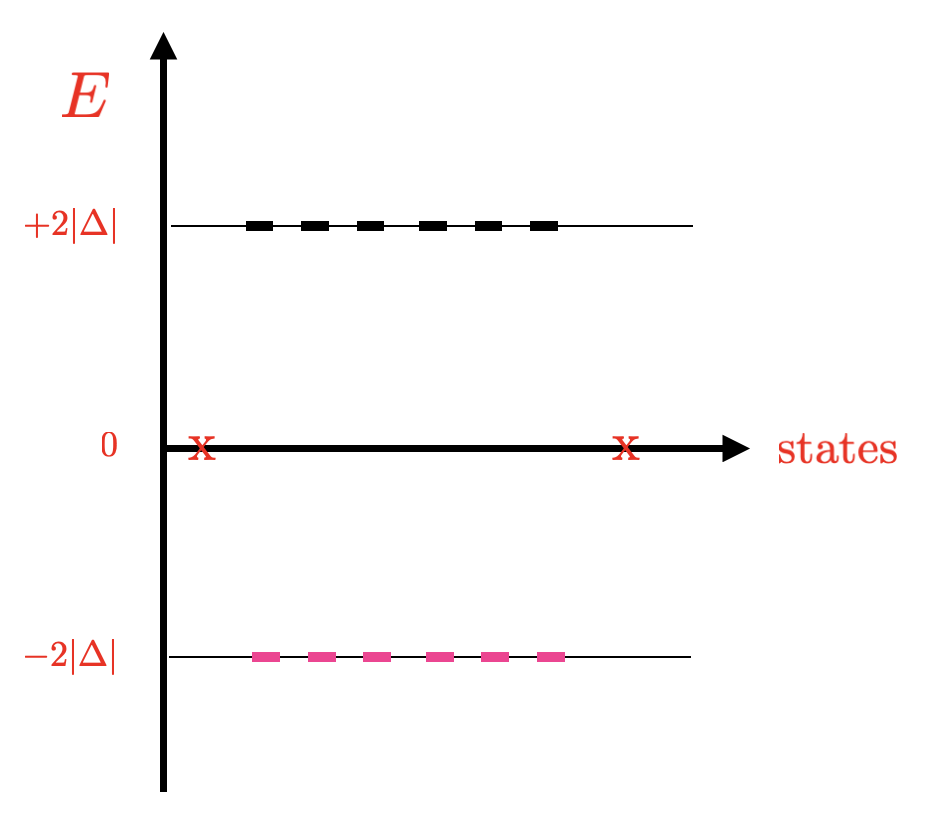}
  \caption{(top)Topological phase of Majorana bound states in a fermionic chain with L=7 sites. Ovals denote sites hosting one single fermion,  each composed of two Majorana modes (solid circles).  In the topological phase, Majorana modes are coupled between neighboring sites effectively breaking the original fermions into parts.  This breaking of the fermionic degrees of freadom is only visible at the end points of the chain where the Majorana bound states manifest unpaired.  (bottom) Energy spectrum of the Hamiltonian \eqref{Kitaev_H} corresponding to the fine tuning of the parameters \eqref{fine_tuning}.}
    \label{fig:unpaired_M}
\end{center}
\end{figure}
%%%%%%
%%%%%%
This Hamiltonian shows no coupling to the end-point Majorana modes $ m_{1} $ and $ m_{2L} $ that have been decoupled from the rest of the modes in the bulk of the chain. These Majorana end-modes thus have zero energy.  In particular,  the unpaired Majorana modes are eigenstates of the Hamiltonian with eigenvalue zero.  Remarkably,  this is an exact result regardless the length of the chain that makes this case very special unlike other cases of Majorana states.  Whereas the modes in the bulk can have either energy $-2|\Delta|$ or $ +2|\Delta| $.  These two values correspond to two flat bands in the energy spectrum (see Fig. \eqref{fig:unpaired_M}).  The ground state corresponds to populating the lowest energy band with particle modes and leaving empty (particle holes)  the highest band.  The two unpaired Majorana modes are located in the middle of the energy gap at zero energy.  This energy configuration is particle-hole invariant.

What is well-defined is the relative energy between the end and bulk states. However, there is a clear energetic difference between those sets of states: the end states are decoupled from the Hamiltonian whereas the bulk states appear as interacting terms in the Hamiltonian. 

Being a zero mode means it has zero energy and zero group velocity.  Whereas the inertial mass $ m $ appearing in \eqref{hopping} has a fundamental origin,   the mass of a quasiparticle zero mode is something effective and has a quantum origin.  It is different from the gravitational mass of a macroscopic body like a planet or the mass of an elementary particle.   It corresponds to a property of the energy band spectrum in k-space (crystal momentum).

Thanks to the SP,  the unpaired Majorana modes are fully decoupled from the bulk modes and in particular,  they show no dependency on the mass trough the hopping parameter $ t $.  These Majorana end-modes are quasiparticles different from the original fermions,  which do depend on the mass.  This unpairing is a consequence of the SP and is topologically protected (parity particle-hole).  Thus, we may say that MBS have zero momentum and energy that is independent of the mass of the bulk.  We may conclude that they have zero energy and zero mass.

Let us remark that a quasiparticle is a subsystem (excitation) of a composite system (many-body ground state). It is the result of a nontrivial superposition of many quantum states akin of quantum matter waves. This subsystem does not have gravitational mass but its effective mass has an energy origin and as such can suffer from gravitational interaction as any other form of energy that is able to gravitate according to Einstein field equations. It contributes to the energy-stress tensor in the field equations (without cosmological constant) \cite{Einstein_relativity,Carroll,Hartle}:
\begin{equation}
G_{\mu\nu} = \kappa T_{\mu\nu}.
\end{equation}
But it so happens that the MBS in 1D has neither energy nor velocity: they are zero modes pinned at the ends of the one-dimensional lattice,  and is a subsystem decoupled from the rest.  It is the bulk what may be subjected to gravitational effects,  despite being negligible in practice, but not the MBS. We may conclude that they are equivalent to a neutral gravitational system.

Notice that for a Majorana particle as a candidate to describe neutrino elementary particles in the Standard Model the situation is different: those Majorana elementary particles have neither zero mass nor zero momentum implying that they are not a neutral gravitational system.

There are two important aspects about the decoupling of the Majorana zero modes:
a) The decoupling implies that the mass of the whole 1D system  is associated to the bulk degrees of freedom that appear in the Hamiltonian through the coupling constant $ t $ that carries mass.  On the contrary,  the MBS disappear from the Hamiltonian and are considered as massless.
b) The decoupling of MBS corresponds to quasiparticles of different nature than the bulk quasiparticles (fermions).  Nevertheless,  the decoupling does not allow them to become free particles because they are not real fermions.

It is interesting to note that there are other types of topological zero edge states at the end of one-dimensional chains but they are not truly massless unless they reach the size of the thermodynamic limit, thereby escaping the gravitationally neutral condition.
Majorana bound states are remarkable in this regard since the fine tuning of coupling parameters in the Hamiltonian \eqref{fine_tuning} produces massless modes at finite lenght and they are localized in real space.

There are several ways out for constructing Majorana states to avoid confrontation with the SEP by circumvating some of their defining properties.  The following is a short representative list of Majorana states that are not gravitationally neutral unlike the previously considered very special case \eqref{fine_tuning}:

\noindent {\em  Massive MBS in 1D.}
Being a zero total energy $E=0$ and zero momentum $p=0$, a Majorana bound state has effectively zero mass $m=0$. This makes it a candidate for a neutral gravitational system. A way to avoid these deadlock properties is by  providing the quasiparticle with an effective mass. In this context this is equivalent to having a gapped excitation with energy gap $\Delta = m$.  Gapped Majorana modes have been proposed in one-dimensional lattices with long-range interactions \cite{long_range,long_rangeJJ}.

\noindent {\em  Massless chiral Majorana boundary states in 2D.}	
Another way out to have Majorana qusiparticles not gravitationally neutral is to provide them with a non-vanishing  momentum $\vec{p}\neq 0$,  while remaining massless.   This may be achieved with models in 2D lattices with the Majorana quasiparticle running through the boundaries of the lattice  \cite{chiral,chiral_longrange}.  	

\noindent {\em  Finite-size Effects in MBS in 1D.}
This possibility arises since the MBS have their properties in the ideal case of fine tuning coupling constants \eqref{fine_tuning}.  Away from this contition,  the MBS receive finite-size contributions.  This may only vanish by  reaching the thermodynamic limit 
$N\rightarrow \infty$, with $N$ the number of sites in the 1D lattice. 	In real samples,  \eqref{fine_tuning} is never really achieved and as a result, the MBS have an effective mass that manifests in the form of a finite correlation length for the pair of Majorana excitations at the end of the chain.  Thus, they cease to be a gravitationally neutral system.

After confronting the properties of the Majorana bound states with the strong equivalence principle, 
we are driven to the following scenarios to provide a framework for the gravitational consequences of the quantum properties of MBS:

\noindent {\em A/ Undeterminable.}
This scenario assumes that the laws of general relativity do not apply to the nano/atomic scale. This possibility describes the situation in which gravity really starts to disappear as we move into the microscopic quantum realm. This would correspond to no real need for a theory of quantum gravity for they would live on different scales and do not interact. Or else, the finite-size effects are unremovable for large lattice sizes.

\noindent {\em B/ Determinable.}
This scenario assumes that the laws of general relativity are applicable at the nano/atomic scale. This scenario corresponds to acknowledging that a theory of quantum gravity is needed one way or another. Then, there are two alternatives in turn:

{\em 1) Broken SEP.}
The SEP is broken by quantum effects and MBS \eqref{fine_tuning} can be constructed.

{\em 2) Unbroken SEP.}
MBS \eqref{fine_tuning} cannot be constructed for they would violate the SEP that remains valid.

%%%%%%%%%%%%%%%%%%%%%%%%%%%%%%%%%%%%%%
%%%%%%%%%%%%%%%%%%%%%%%%%%%%%%%%%%%%%%
\section{Conclusions}
\label{sec:conclusions}
%%%%%%%%%%%%%%%%%%%%%%%%%%%%%%%%%%%%%%
%%%%%%%%%%%%%%%%%%%%%%%%%%%%%%%%%%%%%%

The possibility of breaking the SEP using quantum effects as a matter of principle has been investigated 
by means of Majorana bound states. 
Black holes are traditionally the labs for quantum gravity gedanken experiments aiming at providing hints as to how to build a theory that somehow includes general relativity and quantum mechanics as limiting cases \cite{hawking74,firewalls,er=epr} (assuming scenario B/ above).  Here I have brought up another possibility to confront both theories: the experimental construction of ideal Majorana zero mode states \eqref{fine_tuning} in 1D could be telling us whether some of the basic principles of those theories are at stage, either the strong equivalence principle or the superposition principle. 

Practical consequences of this fact are difficult to assess since once again gravitation is so feeble at the quantum scale that has no consequences.   Possible ways to manifest consequences of this violation would imply the coupling of the MBS to some external field,  but they are chargeless and gravitationless and it seems not a straightforward way to do it.

Assuming that Majorana zero modes are neutral gravitational systems as in the scenario B/1),   what consequences could be derived? Notice that a gravitationally neutral system in particular does not feel gravitation.  Thus,  systems like those would escape the action of black holes: if they were outside the event horizon would never fall in or if they were inside,   nothing could prevent them from escaping.  In the special case of the MBS \eqref{fine_tuning}, there is a caveat though: they do not exist alone but tied up to the bulk of the chain.  Then,  the bulk carries the energy gap and makes it a gravitating object that causes the whole system to gravitate and cannot escape black holes either.  Thus, it is possible to find gravitationally neutral systems that can not escape a black hole by resorting to quantum mechanics.

Thus, we may say that the SEP is violated locally at the ends of the chain but not globally as a whole.   A similar situation happens in topological insulators showing local violations of the second law of thermodynamics (SLT) at the boundaries \cite{topo_heat_I, topo_heat_II,topo_heat_III}.  The big difference is that the SLT has a statistical origin that allows those local violations whereas the SEP, being deterministic,  does not.

If the breaking of SEP is meaningful within the scenario B/1),   then it is an additional incentive to search for MBS. In fact, even more challenging than their use in topological quantum computation \cite{non-abelian,rmp} since the fine tuning condition \eqref{fine_tuning}  is strictly necessary to ensure the massless condition without finite size effects leading to a gravitationally neutral system.  On the contrary, if the scenario B/2) is the one favored by nature, then it would provide an explanation to the difficulties encountered so far in experiments to sinthesize  MBS in the lab \cite{passing_protocol_I,passing_protocol_II}.

\acknowledgements
Support is acknowledged from the CAM/FEDER Project No.S2018/TCS-4342 (QUITEMAD-CM), Spanish MINECO grants MINECO/FEDER Projects,  PID2021-122547NB-I00 FIS2021,  MCIN with funding from European Union NextGenerationEU  PR47/21 MADQuantum-CM PRTR-CM(PRTR-C17.I1) and Ministry of Economic Affairs Quantum ENIA project.  M. A. M.-D. has been partially supported by the U.S.Army Research Office through Grant No. W911NF-14-1-0103.

\appendix

%%%%%%%%%%%%%%%%%%%%%%%%%%%%%%%%%%%%%%%%%%%%%%%%%%%%%%%%%%%%%%%%%%%%%%%%%%%%%%

\end{document}